\documentclass[aps,prl,twocolumn,amssymb,nofootinbib]{revtex4}
\usepackage{graphicx}
\usepackage{amsmath}
\usepackage{amssymb}
\usepackage{bm}

\usepackage{epsfig}
\usepackage{bbm}
\usepackage{color}
\usepackage{ulem}

\definecolor{MyDarkBlue}{rgb}{0.15,0.15,0.45}
\usepackage[linktocpage=true]{hyperref}
\hypersetup{
colorlinks=true,
citecolor=MyDarkBlue,
linkcolor=MyDarkBlue,
urlcolor=MyDarkBlue,
pdfauthor={Kurt Hinterbichler, James Stokes and Mark Trodden},
pdftitle={Cosmologies of extended massive gravity},
pdfsubject={hep-th}
}

\usepackage{enumitem}
\usepackage{graphicx}
\usepackage{subfigure}
\usepackage{amsmath}
\usepackage{bbm}
\usepackage{color}

\newcommand{\be}{\begin{equation}}
\newcommand{\ee}{\end{equation}}
\newcommand{\bea}{\begin{eqnarray}}
\newcommand{\eea}{\end{eqnarray}}
\newcommand{\skappa}{\sqrt{-\kappa}}

\newcommand{\nn}{\nonumber}

\begin{document}

\title{Cosmologies of extended massive gravity} 
\author{Kurt Hinterbichler}
\author{James Stokes}
\author{Mark Trodden}

\affiliation{Center for Particle Cosmology, Department of Physics and Astronomy, University of Pennsylvania,
Philadelphia, Pennsylvania 19104, USA}
\date{\today}

\begin{abstract}
We study the background cosmology of two extensions of dRGT massive gravity.  The first is variable mass massive gravity, where the fixed graviton mass of dRGT is replaced by the expectation value of a scalar field.  We ask whether self-inflation can be driven by the self-accelerated branch of this theory, and we find that, while such solutions can exist for a short period, they cannot be sustained for a cosmologically useful time.  Furthermore, we demonstrate that there generally exist future curvature singularities of the ``big brake" form in cosmological solutions to these theories. The second extension is the covariant coupling of galileons to massive gravity.  We find that, as in pure dRGT gravity, flat FRW solutions do not exist.  Open FRW solutions do exist -- they consist of a branch of self-accelerating solutions that are identical to those of dRGT, and a new second branch of solutions which do not appear in dRGT.

\end{abstract}

\maketitle

\section{Introduction and Outline}

An interacting theory of a massive graviton, free of the Boulware-Deser mode \cite{Boulware:1973my}, has recently been discovered \cite{deRham:2010ik,deRham:2010kj} (the dRGT theory, see \cite{Hinterbichler:2011tt} for a review), allowing for the possibility of addressing questions of interest in cosmology. 
Pure dRGT massive gravity admits self-accelerating solutions \cite{deRham:2010tw,Koyama:2011xz,Nieuwenhuizen:2011sq,Chamseddine:2011bu,D'Amico:2011jj,Gumrukcuoglu:2011ew,Berezhiani:2011mt}, in which the de Sitter Hubble factor is of order the mass of the graviton. Since having a light graviton is technically natural \cite{ArkaniHamed:2002sp,deRham:2012ew}, such a solution is of great interest in the late-time universe to account for cosmic acceleration. 

A natural question is whether a similar phenomenon might drive inflation in the early universe.  To use the self-accelerating solution of massive gravity for inflation (i.e. ``self-inflation''), the graviton mass would have to be of order the Hubble scale during inflation.  Yet, we know that the current graviton mass cannot be much larger than the Hubble scale today \cite{Goldhaber:2008xy}.  

Thus, for self-inflation to be possible, the graviton mass must change in time.  One idea of how to realize this is to promote the graviton mass to a scalar field, $\Phi$, which has its own dynamics and can roll \cite{D'Amico:2011jj,Huang:2012pe}.  The expectation value (VEV) of $\Phi$ then sets the mass of the graviton.  We can imagine that at early times $\Phi$ has a large VEV, so that the graviton is very massive, and the universe self-inflates with a large Hubble constant.  Then, at late times, $\Phi$ rolls to a smaller VEV, self-inflation ends and the graviton mass attains a small value consistent with present day measurements.

In this letter, we will see that, in practice, such an inflation-like implementation of massive gravity is difficult to achieve in this model.  Pure dRGT theory has a constraint, stemming from the Bianchi identity, which forbids standard FRW evolution in the flat slicing \cite{D'Amico:2011jj} (the self-accelerating solutions are found in other slicings).  There appears an analogous constraint in the variable mass theory, and this constraint, while it no longer forbids flat FRW solutions, implies that self-inflation cannot be sustained for a cosmologically relevant length of time. In addition, we show that non-inflationary cosmological solutions to this theory may exhibit future curvature singularities of the ``big brake" type.

In the second half of this letter (which can be read independently from the first), we consider the covariant galileon extension of massive gravity introduced in \cite{Gabadadze:2012tr}. This theory has a new scalar degree of freedom $\pi$, which describes brane bending in an additional spatial dimension.  (Unlike the variable mass theory, $\pi$ here does not act to set the graviton mass, which is fixed.  So we are not interested in self-inflation here, but are just exploring the basic cosmological equations.)  

We derive the background cosmological equations for this theory, and find that the presence of the scalar leads to a more complicated constraint than in pure dRGT.  We discuss the possible solutions in the case of zero and negative spatial curvature.   We find that, as in pure dRGT theory, this constraint forbids flat FRW solutions.   For an open FRW ansatz, however, solutions can exist and they come in two branches. The first branch consists of self-accelerating solutions that are identical to the self-accelerating solutions of pure dRGT theory.  The second branch consists of novel solutions which are not found in pure massive gravity.

\section{Variable Mass Massive Gravity}

We start with variable mass massive gravity.  This is dRGT theory in which the graviton mass squared is promoted to a scalar field $\Phi$,

\be
S=S_{\rm EH}+S_{\rm mass}+S_{\Phi} \ ,
\ee
where
\bea
S_{\rm EH} &=& \frac{1}{2} M_{\rm P}^2\int {\rm d}^4 x \, \sqrt{-g}\, R \ ,  \\
S_{\rm mass} &=& M_{\rm P}^2\int {\rm d}^4 x \, \sqrt{-g}\, \Phi\, (\mathcal{L}_2 + \alpha_3\mathcal{L}_3 + \alpha_4\mathcal{L}_4) \ , \\
S_{\Phi} &=& -\int {\rm d}^4 x \, \sqrt{-g}\, \left[\frac{1}{2}g(\Phi)(\partial \Phi)^2 +V(\Phi)\right] \ . 
\eea
Here $\alpha_3,\alpha_4$ are the two free parameters of dRGT theory.  We have allowed for an arbitrary kinetic function $g(\Phi)$ and potential $V(\Phi)$, so that there is no loss of generality in the scalar sector.
The mass term consists of the ghost-free combinations \cite{deRham:2010kj},
\begin{eqnarray}
 {\cal L}_2 & = & \frac{1}{2}
  \left([{\cal K}]^2-[{\cal K}^2]\right)\,, \nonumber\\
 {\cal L}_3 & = & \frac{1}{6}
  \left([{\cal K}]^3-3[{\cal K}][{\cal K}^2]+2[{\cal K}^3]\right), 
  \nonumber\\
 {\cal L}_4 & = & \frac{1}{24}
  \left([{\cal K}]^4-6[{\cal K}]^2[{\cal K}^2]+3[{\cal K}^2]^2
   +8[{\cal K}][{\cal K}^3]-6[{\cal K}^4]\right)\,, \nn \\
\end{eqnarray}
where $K^\mu_{\ \nu} = \delta^\mu_{\ \nu} - \sqrt{g^{\mu\sigma}\eta_{\sigma\nu}}$, $\eta_{\mu\nu}$ is the non-dynamical fiducial metric which we have taken to be Minkowski, and the square brackets are traces.  To work in the gauge invariant formalism, we introduce  four St\"uckelberg fields $\phi^a$ through the replacement $\eta_{\mu\nu}\rightarrow \partial_\mu\phi^a\partial_\mu\phi^b\eta_{ab}$.

Variable mass massive gravity was first considered in \cite{D'Amico:2011jj}, and further studied in \cite{Huang:2012pe,Saridakis:2012jy,Cai:2012ag,newpaper} (see also \cite{D'Amico:2012zv} for a more symmetric scalar extension of dRGT).
dRGT gravity has been demonstrated to be ghost-free through a variety of different approaches \cite{Hassan:2011hr,Hassan:2011ea,Mirbabayi:2011aa,deRham:2011rn,Hinterbichler:2012cn}, and the introduction of the scalar field does not introduce any new Boulware-Deser like ghost degrees of freedom into the system \cite{Huang:2012pe}. 

For cosmological applications we take a Friedmann, Robertson-Walker (FRW) ansatz for the metric, so that
\begin{equation}\label{FRWans}
	\mathrm{d} s^2 = - N^2(t)\mathrm{d}t^2 + a^2(t)\Omega_{ij} \mathrm{d} x^i \mathrm{d}x^j,
\end{equation}
where 
\begin{equation}
	\Omega_{ij} = \delta_{ij} + \frac{\kappa}{1-\kappa r^2}x^i x^j
\end{equation}
is the line element for a maximally symmetric 3-space of curvature $\kappa$ and $r^2 = x^2+y^2+z^2$.  We also take the assumptions of homogeneity and isotropy for the scalar field,
\begin{equation}\label{FRWansscalar}
\Phi=\Phi(t).
\end{equation}

Consider first the case of flat Euclidean sections ($\kappa=0$). We work in the gauge invariant formulation, and the Stueckelberg degrees of freedom take the ansatz \cite{D'Amico:2011jj,Gumrukcuoglu:2011ew}.
\be \phi^i = x^i, \ \  \phi^0 = f(t),\label{stukans}\ee
 where $f(t)$, like $a(t)$, is a monotonically increasing function of $t$.

Inserting \eqref{FRWans} and \eqref{stukans} into the action, we obtain the mini-superspace action

\begin{align}
	S_{\rm EH} 
		& = 3 M_{\rm P}^2 \int{\rm d}t\, \left[ - \frac{\dot{a}^2a}{N}\right] ,\\
	S_{\rm mass}
		& = 3 M_{\rm P}^2 \int{\rm d}t\,  \Phi\left[NF(a) - \dot{f}G(a)\right],\\
	S_\Phi
		& = \int{\rm d}t \, a^3\left[ \frac{1}{2} N^{-1}g(\Phi)\dot{\Phi}^2 - NV(\Phi) \right] .
\end{align}
where
\begin{align} \nn
	F(a)
		& = a(a-1)(2a-1) \\ & \quad + \frac{\alpha_3}{3}(a-1)^2(4a-1) + \frac{\alpha_4}{3}(a-1)^3, \label{Fofa}
 \\
	G(a)
		& = a^2(a-1)+\alpha_3 a(a-1)^2 + \frac{\alpha_4}{3}(a-1)^3.\label{Gofa}
\end{align}
This mini-superspace action is invariant under time reparametrizations, under which $f$ transforms like a scalar.

There are four equations of motion, obtained by varying with respect to $F,N,\Phi$ and $a$.  As in GR, the Noether identity for time reparametrization invariance tells us that the acceleration equation obtained by varying with respect to $a$ is a consequence of the other equations, so we may ignore it.  After deriving the equations, we will fix the gauge $N=1$ (this cannot be done directly in the action without losing equations).

Varying with respect to $f$ we obtain the constraint pointed out in \cite{Saridakis:2012jy},
\begin{equation}\label{e:mass}
	\Phi = \frac{\mathcal{C}}{G(a)},
\end{equation}
where $\mathcal{C}$ is an arbitrary integration constant.  (Note that the analogous equation in the fixed mass theory implies that $a={\rm constant}$, so there are no evolving flat FRW solutions in that case \cite{D'Amico:2011jj}.) Varying with respect to $N$, we obtain the Friedmann equation, 
\begin{equation} \label{e:FriedmannFlat}
3 M_{\rm P}^2 \left[ H^2 + \frac{\Phi F(a)}{a^3} \right] = \frac{1}{2}g(\Phi)\dot{\Phi}^2 + V,
\end{equation}
and varying with respect to $\Phi$ we obtain the scalar field equation
\begin{align}\nn
	  g(\Phi)\left[{\ddot \Phi} +3 H {\dot \Phi}\right] & +{1\over 2}g'(\Phi)\dot\Phi^2+V'(\Phi)
\\ \quad &= 3M_{\rm P}^2\left[\frac{F(a)}{a^3} - {\dot f}\frac{G(a)}{a^3}\right].
\label{e:phieomFlat}
\end{align}

Rather than solving the coupled second-order Einstein-scalar equations of motion, one can instead reduce the system to a single first-order Friedmann equation.
The relation \eqref{e:mass} can be used to eliminate $\Phi$ and its first derivative from \eqref{e:FriedmannFlat}, which then becomes a first-order differential equation in $a$ which determines the scale factor,
\begin{equation}
	H^2
		=\frac{V\left({\cal C}\over G(a)\right)-3M_P^2{\cal C}{F(a)\over a^3 G(a)}}{3M_P^2-{1\over 2}{\cal C}^2g\left({{\cal C}\over G(a)}\right){G'(a)^2a^2\over G(a)^4}}.\label{modfried}
\end{equation}
Once we have solved for the scale factor, the scalar $\Phi$ is determined from  \eqref{e:mass} and the Stueckelberg field $f$ is determined by  solving \eqref{e:phieomFlat}\footnote{Note that in general the Stueckelberg field cannot be chosen arbitrarily as in \cite{Saridakis:2012jy} but is non-trivially constrained by the choice of mass term, or in this case, kinetic function $g(\Phi)$.}.

\subsection{Singularities}

One feature of this model that has not been noticed previously is that it allows for the possibility of curvature singularities at finite values of $a$. These can happen when the evolution attempts to pass through values of $a$ for which the denominator of the right hand side of \eqref{modfried} goes to zero.  

When this happens $a$ is finite, but the Hubble parameter, and hence $\dot a$, is blowing up.  The scalar curvature is also blowing up, so this is a genuine curvature singularity; a ``big brake'' where the universe comes to some finite scale factor and gets stuck \cite{Barrow:2004xh,Dabrowski:2007dn}.  Similar types of singularities also occur in DGP \cite{Gregory:2007xy}.  

For example, Taylor expanding the denominator of \eqref{modfried} for large $a$ we obtain a critical value of the scale factor at which the Hubble parameter diverges,
\begin{equation}
	a_{\rm cr} = \sqrt{3}\left(\frac{g(0)}{2 M_{\rm P}^2}\right)^{1/6}\left(\frac{\mathcal{C}}{3+3\alpha_3 + \alpha_4}\right)^{1/3}.
\end{equation}

\subsection{Self-Inflation}

We now consider the possibility that the graviton has a large mass in
the early universe, through some natural displacement (and resulting VEV) of the scalar field from its true minimum near zero.
We seek dynamics such that the scalar field slowly rolls down its potential, during which time the graviton remains massive, resulting in a large self-acceleration of the universe.  This self-acceleration comes from the second term on the left-hand side of the Friedmann equation \eqref{e:FriedmannFlat}.  For this to be true self-acceleration this term should be much larger than both the scalar kinetic energy and potential on the right hand side, so that the acceleration is primarily driven by the modification to gravity and not by the scalar field. 
After many e-folds, $\Phi$ should roll towards zero, self-inflation should end, and the graviton mass should become small at late times.

Thus, assume we have an inflationary solution $a\sim e^{Ht}$, with $H\sim {\rm constant}$. 
The scale factor is growing exponentially, so we Taylor expand the entire right hand side of \eqref{modfried} for large $a$, as
\be H^2={V(0)\over 3M_P^2}+{\cal C}\left[{V'(0)\over M_P^2}-(6+4\alpha_3+\alpha_4)\over 3+3\alpha_3+\alpha_4\right]{1\over a^3}+{\cal O}\left(1\over a^4\right).
\ee
We see that the dependence on all of the massive gravity modifications redshifts away exponentially, at least as fast as ${a^{-3}}$, and we are left with inflation driven only by the value of the potential at $\Phi=0$.  (In particular, contributions sensitive to the function $g$ only start to enter at ${\cal O}(1/a^7)$.) 

Said another way, we only have self-inflation if the quantity $\frac{\phi F(a)}{a^3}$ in \eqref{e:FriedmannFlat} is approximately constant when $a\sim e^{Ht}$.  But the constraint equation \eqref{e:mass} makes this impossible: we see from \eqref{e:mass} that $\Phi\sim {1\over a^3}$, since $G(a)\sim a^3$ for large $a$.  Since $F(a)\sim a^3$, the quantity $\frac{\Phi F(a)}{a^3}$ behaves like $\sim \Phi\sim { a^{-3}}$, so it goes to zero exponentially fast and we cannot sustain self-inflation.

Having encountered an obstacle to the possibility of self-inflation in the flat slicing, we now investigate the possibility in the open slicing ($\kappa < 0$). Following \cite{Gumrukcuoglu:2011ew} we take the Stueckelberg ansatz to be
\begin{align}\label{curvansatz}
	\phi^0
		& = f(t) \sqrt{1-\kappa r^2} ,\ \ \phi^i
		 = \sqrt{-\kappa} f(t) x^i.
\end{align}
The mini-superspace action then becomes
\begin{align}
	S_{\rm EH} 
		& = 3 M_{\rm P}^2 \int {\rm d}t \,\left[ - \frac{\dot{a}^2a}{N} + \kappa N a\right], \\
	S_{\rm mass}
		& = 3 M_{\rm P}^2 \int {\rm d}t\, \Phi \left[NF(a,f) - \dot{f}G(a,f)\right], \\
	S_\Phi
		& = \int {\rm d}t\, a^3\left[ \frac{1}{2} N^{-1}g(\Phi)\dot{\Phi}^2 - NV(\Phi) \right] ,
\end{align}
where
\begin{align}\nn
	F(a,f)
		& = a(a-\sqrt{-\kappa}f)(2a-\sqrt{-\kappa}f)  \\  \label{Fofaf} 
		& +  \frac{\alpha_3}{3}(a-\sqrt{-\kappa}f)^2(4a-\sqrt{-\kappa}f) + \frac{\alpha_4}{3}(a-\sqrt{-\kappa}f)^3, \\ \nn
	G(a,f)
		& = a^2(a-\sqrt{-\kappa}f)+\alpha_3 a(a-\sqrt{-\kappa}f)^2  \\ &+ \frac{\alpha_4}{3}(a-\sqrt{-\kappa}f)^3. \label{Gofaf}
\end{align}
Again, we have time reparametrization invariance so we can ignore the acceleration equation, and we will fix the gauge $N=1$ after deriving the equations of motion.  The constraint equation arising from varying with respect to $f$ is
\be
({\dot a}-\skappa)\Phi\frac{\partial G(a,f)}{\partial a}+ G(a,f){\dot \Phi}=0 \ .
\label{fconstraintk}
\ee

The Friedmann equation obtained by varying with respect to $N$ is
\be
3 M_{\rm P}^2 \left[ H^2 + \frac{\kappa}{a^2}  +\frac{\Phi F(a,f)}{a^3} \right] = \frac{1}{2}g(\Phi)\dot{\phi}^2 + V(\Phi),
\label{e:FriedmannCurved}
\ee
and the equation of motion for $\Phi$ is 
\begin{align}\nn
	   g(\Phi)\left[{\ddot \Phi}+3 H {\dot \Phi}\right]&+{1\over 2}g'(\Phi)\dot\Phi^2+V'(\Phi)   \\
		& \quad \ \ \ \ = 3M_{\rm P}^2\left[\frac{F(a,f)}{a^3} - {\dot f}\frac{G(a,f)}{a^3}\right].
\label{e:phieomk}
\end{align}

In order to obtain inflation driven by the graviton mass, the term $\Phi F(a,f)/a^3$ in the Friedmann equation \eqref{e:FriedmannCurved} must be approximately constant when $a\sim e^{Ht}$.  Rearranging the constraint equation \eqref{fconstraintk} to isolate $\Phi$ gives
\be {\dot \Phi\over \Phi}=-\left(H-{\sqrt{-\kappa}\over a}\right){a {\partial G(a,f)\over \partial a}\over G(a,f)}.\label{phidk}\ee

Now there are three possibilities: the first is that $a(t)\sim e^{Ht}$ grows faster than $f(t)$.  In this case, the right-hand side of \eqref{phidk} approaches a constant at late times, $\dot{\Phi}/\Phi \sim -3H+{\cal O}(1/a)$, which tells us that $\Phi$ decreases exponentially at late times, $\Phi(t)\sim e^{-3Ht}$.  This, in turn, implies that the self-acceleration quantity $\Phi F(a,f)/a^3$ in \eqref{e:FriedmannCurved} decreases exponentially like $\Phi$, since $F(a,f)/a^3$ approaches a constant.  So once again, we cannot sustain inflation in this case. The second possibility is that $f(t)$ grows faster than $a(t)\sim e^{Ht}$.  In this case, the right-hand side of \eqref{phidk} goes to zero at late times, so $\Phi$ becomes constant.  The self-acceleration quantity $\Phi F(a,f)/a^3$ in \eqref{e:FriedmannCurved} now grows without bound as $\sim f^3(-\kappa)^{3/2}/a^3$ at late times, so again we do not achieve sustained inflation.  Finally, there is the possibility that $f(t)\sim e^{-Ht}$, growing at the same rate as $a(t)$.   This case follows the same pattern as the first possibility -- the right-hand side of \eqref{phidk} approaches a constant at late times, and the self-acceleration quantity decreases exponentially.

In summary, flat and open FRW solutions exist in mass-varying massive gravity, but the constraint equation (the one which forbids flat FRW solutions in pure dRGT theory) does not allow for long-lasting self-inflation.   Finally, homogeneous and isotropic closed FRW solutions are not possible for the same reason they are not in dRGT -- the fiducial Minkowski metric cannot be foliated by closed slices.

\section{The DBI Model}

We now turn to the cosmology of the coupled galileon-massive gravity model introduced in \cite{Gabadadze:2012tr}. One way to construct the theory is to consider a dynamical metric, $g_{\mu\nu}$, on the worldvolume of a 3-brane in a five dimensional flat bulk, with embedding coordinates $X^A(x)$, coupled to the induced metric through dRGT interactions. The worldvolume theory is
\begin{equation}
	S[g,\bar{g}] = S_{R}[g] + S_{\rm mass}[g,\bar{g}] + S_\pi[\bar{g}],
\end{equation}
where $S_R$ and $S_{\rm mass}$ are the usual Einstein-Hilbert and dRGT mass terms except that now 
\be \bar{g}_{\mu\nu}=\partial_\mu X^A\partial_\nu X^B\eta_{AB}\ee is the pull-back of the 5D Minkowski metric to the 4D brane.  The action $S_\pi$ consists of ghost-free scalars constructed from $\bar{g}_{\mu\nu}$ and the extrinsic curvature \cite{deRham:2010eu}.  The only term which will be relevant for us is the DBI term
\be S_\pi  = -\Lambda^4 \int{\rm d}^4 x \sqrt{-\bar g},\ee
where $\Lambda$ is some characteristic mass scale.  In the flat slicing, the other possible higher order terms in $S_\pi$ all vanish on the cosmological backgrounds we consider, so we may exclude them without loss of generality.  They do contribute in the curved slicing, however the final constraint equation \eqref{DBIconstraintk} and the conclusions drawn from it are not modified in their presence, so we ignore them in what follows.
The embedding functions are scalar fields; the first four become the Stueckelberg fields, and the fifth becomes an extra physical scalar, the brane bending mode $\pi$,
\be  X^\mu=\phi^\mu,\ \ \ X^5\equiv \pi.\ee
 
Consider first the case of flat Euclidean space ($\kappa=0$).  We take the FRW ansatz \eqref{FRWans} and \eqref{stukans}, along with homogeneity for the scalar $\pi=\pi(t)$.  The mini-superspace action is

\begin{align}
	S_R 
		& = 3 M_{\rm P}^2 \int{\rm d}t \, \left[ - \frac{\dot{a}^2a}{N}\right] ,\\
	S_{\rm mass}
		& = 3 M_{\rm P}^2 \int{\rm d}t \,  m^2 \left[NF(a) - G(a)\sqrt{\dot{f}^2-\dot{\pi}^2} \right],\\
	S_\pi 
		& = -\Lambda^4 \int{\rm d}t \, \sqrt{\dot{f}^2-\dot{\pi}^2},
\end{align}
where $F(a)$ and $G(a)$ are as in \eqref{Fofa} and \eqref{Gofa}.

The mini-superspace action is invariant under time reparametrizations, under which both $f$ and $\pi$ transform like scalars.  Thus, as before we may ignore the acceleration equation of motion obtained by varying with respect to $a$, since it is redundant.  We fix the gauge $N=1$ after deriving the equations.

Varying with respect to $N$ we obtain the Friedmann equation
\begin{equation}
	H^2 + m^2 {F(a)\over a^3} = 0.
\end{equation}
The $f$ and $\pi$ equations are, respectively
\begin{equation}\label{DBIfequation}
	{\delta S\over \delta f}={d\over dt}\left[\left(3M_{\rm P}^2 m^2G(a) + \Lambda^4 \right)\frac{\dot{f}}{\sqrt{\dot{f}^2-\dot{\pi}^2}} \right]= 0,
\end{equation}
\begin{equation}\label{DBIpiequation}
	{\delta S\over \delta \pi}=-{d\over dt}\left[\left(3M_{\rm P}^2 m^2G(a) +\Lambda^4 \right)\frac{\dot{\pi}}{\sqrt{\dot{f}^2-\dot{\pi}^2}} \right]= 0.
\end{equation}

Taking the combination $\dot f{\delta S\over \delta f}+\dot \pi{\delta S\over \delta \pi}$ of the two equations of motion \eqref{DBIfequation} and \eqref{DBIpiequation} , we arrive at the following constraint equation,
\begin{equation}\label{DBIconstraint}
	\sqrt{\dot{f}^2-\dot{\pi}^2}{d\over dt}\left[G(a)\right] = 0.
\end{equation}

The square root cannot be zero because it appears in the denominator of  \eqref{DBIfequation} and \eqref{DBIpiequation}, so we must have ${d\over dt}\left[G(a)\right]=0$, which implies that $a$ must be constant and thus there are no evolving flat FRW solutions.  This is the same phenomenon as in pure dRGT theory \cite{D'Amico:2011jj}, and we see that the DBI extended theory has a similar structure.
The inclusion of matter minimally coupled to $g_{\mu\nu}$ does not affect this conclusion because it does not contribute to the equations \eqref{DBIfequation}, \eqref{DBIpiequation} and hence the constraint \eqref{DBIconstraint} is unchanged.

We now turn to the case of nonzero spatial curvature $\kappa < 0$.  Using again the ansatz \eqref{curvansatz} we obtain
\begin{align}
S_R 
		& = 3 M_{\rm P}^2 \int {\rm d}t\,\left[ - \frac{\dot{a}^2a}{N} + \kappa N a\right] ,\\
	S_{\rm mass}
		& = 3 M_{\rm P}^2 \int{\rm d}t \, m^2 \left[NF(a,f) - G(a,f)\sqrt{\dot{f}^2 - \dot{\pi}^2} \right], \\
	S_\pi
		& = -\Lambda^4 \int{\rm d}t \, (\sqrt{-\kappa} f)^3\sqrt{\dot{f}^2-\dot{\pi}^2},
\end{align}
with $F(a,f)$ and $G(a,f)$ as in \eqref{Fofaf} and \eqref{Gofaf}.

Varying with respect to $N$ and setting $N=1$, we obtain the Friedmann equation
\begin{equation}\label{DBINk}
	H^2 + {\kappa \over a^2} + m^2 {F(a,f)\over a^3}  = 0.
\end{equation}

The $f$ and $\pi$ equations are, respectively
\begin{align}\label{DBIfequationk}
	{\delta S\over \delta f} & = 3M_{\rm P}^2 m^2\left[{\partial F(a,f)\over \partial f}-{\partial G(a,f)\over \partial f}\sqrt{\dot{f}^2-\dot{\pi}^2}\right] \nonumber\\
	& -3\Lambda^4(\sqrt{-\kappa})^3 f^2\sqrt{\dot{f}^2-\dot{\pi}^2} + \nonumber \\
	&+{d\over dt}\left[\left(3M_{\rm P}^2 m^2G(a,f) + \Lambda^4(\sqrt{-\kappa}f)^3 \right)\frac{\dot{f}}{\sqrt{\dot{f}^2-\dot{\pi}^2}} \right]= 0,
\end{align}

\begin{align}\label{DBIpiequationk}
&	{\delta S\over \delta \pi}=\nn \\ & -{d\over dt}\left[\left(3M_{\rm P}^2 m^2G(a,f) +\Lambda^4(\sqrt{-\kappa}f)^3 \right)\frac{\dot{\pi}}{\sqrt{\dot{f}^2-\dot{\pi}^2}} \right]= 0.
\end{align}

Taking the combination $\dot f{\delta S\over \delta f}+\dot \pi{\delta S\over \delta \pi}$ of the two equations of motion \eqref{DBIfequationk} and \eqref{DBIpiequationk}, and using the relation ${\partial F(a,f)\over \partial f}=-\sqrt{-\kappa}{\partial G(a,f)\over \partial a}$, which can be checked straightforwardly from the definitions  \eqref{Fofaf} and \eqref{Gofaf}, we arrive at the following constraint equation,
\begin{equation}\label{DBIconstraintk}
	{\partial G(a,f)\over \partial a}\left(\dot a\sqrt{\dot f^2-\dot\pi^2}-\sqrt{-\kappa}\dot f\right)= 0.
\end{equation}

There are now two possible branches of solutions.  The first consists of the solutions for which ${\partial G\over\partial a}=0$.  Solving this algebraically for $f$, we find $f={\cal C_\pm}{a R}$ for some constant ${\cal C_\pm}$ depending only on the coefficients $\alpha_2,\alpha_3$ (there can be two branches  here, since we must solve a quadratic equation for $f$).  Then, reinserting this into \eqref{DBINk}, we obtain a modified Friedmann equation.  The modification $\frac{F(a,{\cal C_\pm}{a R})}{a^3}$ is a constant, depending only on $\alpha_2,\alpha_3$, so that when this constant is negative, we have self-acceleration with $\rho_{\rm self}\sim M_P^2 m^2$.  These solutions are exactly the same self-accelerated solutions that exist for the pure dRGT theory \cite{Gumrukcuoglu:2011ew}.  The solution for $\pi$ can then be determined by solving \eqref{DBIpiequationk}.

However, for the theory at hand, there exists a new possibility. This second branch consists of solutions for which
\begin{equation}\label{branch2}
	\dot a\sqrt{\dot{f}^2-\dot{\pi}^2}=\sqrt{-\kappa}\dot{f}.
\end{equation}
In the case of the pure dRGT theory where $\pi=0$, this branch gives only solutions for which $a=\sqrt{-\kappa}t$, which is just Minkowski space in Milne coordinates.  Here we have the possibility of non-trivial solutions on this branch.
Solving \eqref{branch2} for $\dot{\pi}$ gives $\dot{\pi} = \pm\dot{f}\sqrt{1+\frac{\kappa}{\dot{a}^2}},$ and
substituting this into the $\pi$ equation of motion \eqref{DBIpiequationk} we see that $\dot{f}$ cancels and we are left with an algebraic equation in $f$, namely
\begin{equation}
	\left(3M_{\rm P}^2 m^2G(a,f) +\Lambda^4(\sqrt{-\kappa}f)^3 \right)\sqrt{\frac{\dot{a}^2+\kappa}{-\kappa}} = \mathcal{C}
\end{equation}
where ${\cal C}$ is the integration constant from integrating \eqref{DBIpiequationk}. This is a cubic equation which can be solved for $f$. Eliminating $f$ from the Friedmann equation \eqref{DBINk} yields a separable equation of motion for the scale factor which can have solutions with non-trivial evolution.

\section{Summary and Conclusions}

In this letter we have explored the cosmology of two different extensions of the dRGT massive gravity theory.  

The first model is variable mass massive gravity, which results from promoting the fixed mass term of the dRGT model to the vacuum expectation value of a dynamical scalar field, as suggested in \cite{D'Amico:2011jj}.  In dRGT theory, there is a constraint equation that forbids non-trivial flat FRW solutions, though self-accelerating open solutions exist.  In the variable mass theory, the form of the constraint is different, and it no longer forbids flat FRW solutions.  The constraint, however, makes it difficult to realize the idea of self-inflation, i.e. using the self-acceleration properties of massive gravity in the early universe to drive inflation. Furthermore, we have demonstrated for the first time that a large class of cosmologies within these models exhibit a future curvature singularity of the ``big brake" form.

The second extension is the coupled galileon-massive gravity scalar-tensor theory of \cite{Gabadadze:2012tr}.  In the flat slicing, the constraint equation takes a similar form to that of the pure dRGT theory, and it similarly rules out non-trivial FRW solutions.  In the open slicing, the theory reproduces the self-accelerating branch discovered in pure massive gravity, and in addition, provides a new branch of evolving cosmological solutions, the detailed general properties of which will be the topic of future work.

{\bf Acknowledgments:}
We would like to thank Gregory Gabadadze, A. Emir G\"umr\"uk\c{c}\"uo\u{g}lu and  Emmanuel Saridakis for helpful comments.
MT is supported in part by the US Department of Energy, and JS and MT are supported in part by NASA ATP grant NNX11AI95G.  Research at Perimeter Institute is supported by the Government of Canada through Industry Canada and by the Province of Ontario through the Ministry of Economic Development and Innovation. The work of KH was made possible in part through the support of a grant from the John Templeton Foundation. The opinions expressed in this publication are those of the authors and do not necessarily reflect the views of the John Templeton Foundation.

\end{document}